# Analysis of the Veracities of Industry Used Software Development Life Cycle Methodologies

AZM Ehtesham Chowdhury, Abhijit Bhowmik, Hasibul Hasan, Md Shamsur Rahim

*Abstract*— Currently, software industries are using different SDLC (software development life cycle) models which are designed for specific purposes. The use of technology is booming in every perspective of life and the software behind the technology plays an enormous role. As the technical complexities are increasing, successful development of software solely depends on the proper management of development processes. So, it is inevitable to introduce improved methodologies in the industry so that modern human centred software applications development can be managed and delivered to the user successfully. So, in this paper, we have explored the facts of different SDLC models and perform their comparative analysis.

## I. INTRODUCTION

In the modern world, technology is playing a vital role in everyday life. To make our life easier, day by day new technologies are invented and developed. To reach the level of human comfort, development of the driver software is getting technically complex. To develop the software, development process in the software industry must be more dynamic and adaptive to deal with the complexities [1][2][3]. From several decades, researchers have proposed several software development life cycle models [4]. A software development life cycle model defines the sequential stages of an entire lifetime of a software product [5]. The model is used to divide the project into several actions. Each of the activities goal to provide the good planning and management of the project. The proper planning and management will allow the development team to deliver the product in time and minimize the development cost [6]. From the 60's several SDLC models have been proposed and applied to achieve the better development situation and economic success. SDLC represents the entire process life based on specification, design, validation, evolution [7]. The SDLC gives the outline for the documentation which is necessary to understand client requirements. SDLC helps to define the budget, schedule of the software project [8]. It also provides the elements to analyze time and cost information. SDLC facilitates the guideline to project manager to organize and planning for the project [9]. SDLC contains a sequence of stages [10] where each of them can be characterized as follows:

A. *Initial Analysis*
- Inquire about the organizational objectives.
- Understanding the problems and how to fit with the organization.
- Interviewing client-side stakeholders (end users).
- Description of cost and benefit.

B. *System Requirements*
- Identify project goals.
- Identify functionalities.

C. *Development*
- Code Writing.

D. *Deployment*
- Initial deployment the project to end users.

E. *Maintainance and Evaluation*
- Continuous evaluation till final deployment.
- Changes from initial software.
- Assessment of the development process

F. *Disposal*
- Final version.
- Archive information about the software.

**AZM Ehtesham Chowdhury**
Dept. of CSE
American International University-Bangladesh (AIUB).
Dhaka, Bangladesh
e-mail: ehtesham@aiub.edu

**Abhijit Bhowmik**
Dept. of CSE
American International University-Bangladesh (AIUB).
Dhaka, Bangladesh
e-mail: abhijit@aiub.edu

**Hasibul Hasan**
Dept. of CSE
American International University-Bangladesh (AIUB).
Dhaka, Bangladesh
e-mail: hasib.hasan@aiub.edu

**Md Shamsur Rahim**
Dept. of CSE
American International University-Bangladesh (AIUB).
Dhaka, Bangladesh
e-mail: shamsur@aiub.edu

- Prevention from disclosure of any sensitive data.
- Disposal activities must ensure new systems.

Following above characteristics, various SDLC models have proposed. Some of the popular SDLC models are: Waterfall, Incremental, Iterative, Spiral, Prototyping, XP, SCRUM, RAD, DSDM and so on. These models have their own effectiveness level based on the project and industry. These models are clustered into two different categories: Agile Process and Plan Driven Process. A typical software development process model has generally several stages. Planning and gathering information about the project, Analyze and define requirements, Design and define the product architecture, development, test the product per requirements and technical perspective, deploy product, maintenance.

Rest of this paper is systematized in following sections: In section II and III, we have described the plan driven and agile SDLC models and their advantage and disadvantages. In section IV, we have provided the comparative analysis of the facts of plan driven models and agile models and future work and conclusion in section V.

## II. PLAN DRIVEN MODEL

The products are planned and progress is calculated based on the plan [5]. It is costly to the immediate adaption of changes in requirements for ongoing projects. A plan driven model is best suited for the large teams and enormously critical products which are hard to scale down [11]. These models are effective when the development environment is stable. Experienced stakeholders are required only at the beginning of the project. The success is dependent on structure and order. But these models are less effective for the dynamic development environment. If the changes in requirements befall recurrently, then it is costly [6]. Basic characteristics of several popular plan driven models are described below based on their use in industries:

### A. Waterfall

The waterfall model is the oldest development life cycle model. The waterfall follows sequential development approach. The model facilitates the early planning stages. This model emphasizes to analyze all the requirements and design of the software before started developing [12] [13]. Here the development stages such as requirement analysis, design, development, testing, maintenance depend on the previous phase like the designing phase will be started after finishing off the requirement analysis. So, the software life flows like water falling from the mountain which we know as waterfall. It is still the mostly used SDLC. This model recognizes milestones and widely used for mature products [36]. The waterfall facilitates with a variance of team members. As all the stages depend on the previous one, so any severe flaw appears on any stage the next stages are going to be stuck and go back to previous stages to update. So, the extra time is elapsed from planned time duration. This model does not ensure the feature versions after product deployment.

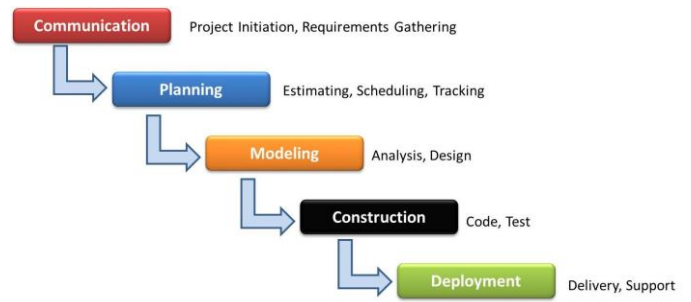

Fig. 1. Waterfall Model [14]

### B. Iterative and Incremental Development Model

The waterfall model does not accommodate any change or feedback from early stages. The iterative model is a class of another plan driven model. The model provides the product to be divided into small parts where each small part known as increments, contains all stages like waterfall [15]. Requirements are also divided and prioritized. On the basis of the requirement set, an increment is defined and highest prioritized requirements are comprised on early increments. This helps the development team to monitor the outcome of the early product. It also allows getting feedback from the system user. More attention and resources are needed to manage the project. The designing issue may occur as all requirements may not analyze earlier.

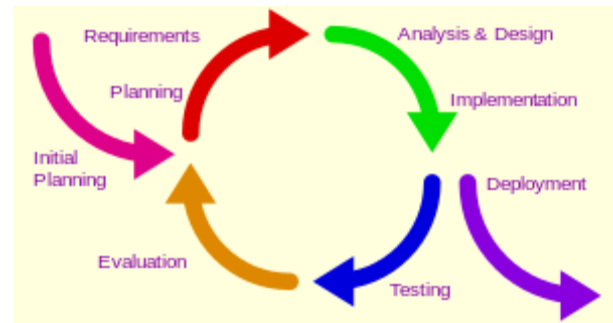

Fig. 2. Iterative and Incremental Development Model [16]

### C. Spiral Development Model

The spiral model is same as the incremental model, but concerns about the risk of the project. Many software development companies are adapting it. It accommodates changes on requirements early as users are involved early in the process. This model visualizes the system or product early [17]. The management of the process is complex like the iterative model as spiral may go an indefinite period and not suitable for small projects.

### D. Prototype Model

The prototype model is mostly suitable when the requirements are not clear utterly. This model aids as a mechanism for recognizing software requirements. It needs user involvement in the early phases more to say before the development phase. It facilitates the modeling of the functionalities of a software in a way that it may not comprise the particular logics of the of the desired software [19]. This model provides well understanding and feedback from the users. This model suffers from insufficient requirement analysis. So, customers may get confused among prototypes and the actual system.

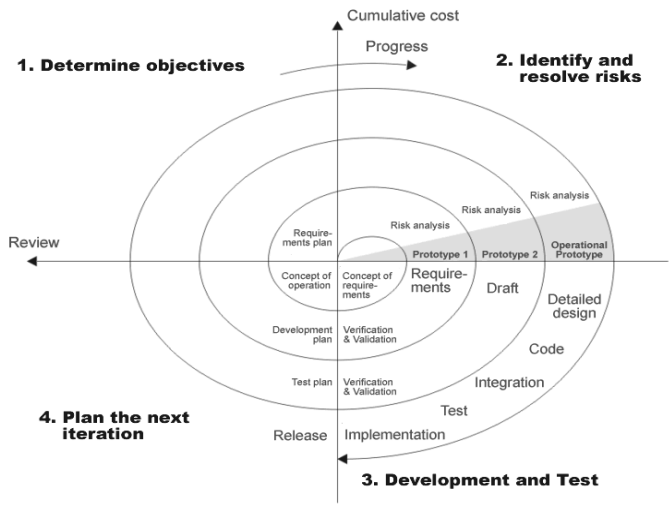

Fig. 3. Spiral Development Model [18]

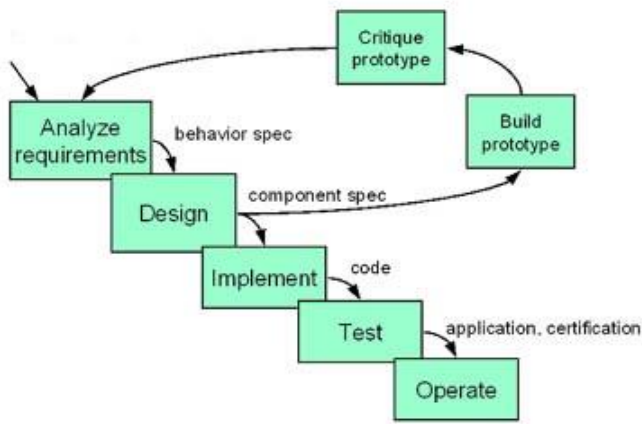

Fig. 4. Prototype Model [20]

### E. *V-Shaped Model*

The V-Shaped life cycle is a sequential processes model like waterfall. Each phase must be completed before initiation of the next phase alike waterfall, but testing is emphasized in this model more than the waterfall model. Before any coding implementation is done, the testing measures are established early in the life cycle during each of the phases. Requirements initiate the life cycle as the waterfall model. A product test plan is established before development has started. The test plan is based on the specified functionalities in requirements collection. The high-level design phase is based on system des-

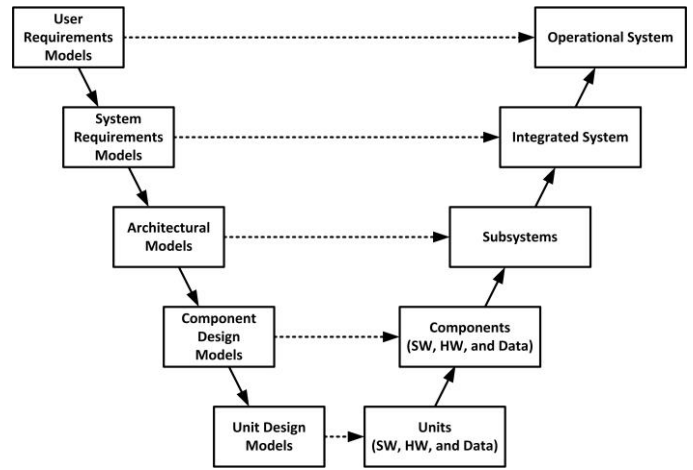

Fig. 5. V- Shaped Model [21]

Ign. The high-level design phase is based on system design. For testing the parts of the products to identify the stability to work together, an integration test plan is fashioned in this stage. The low-level design phase is based on actual software components design. Unit tests are established in this stage. Then all coding is started. After completing the coding, execution path is continued through the right side of the V where the earlier developed test plans are placed to testing [4].

### F. *Rapid Application Development Model*

Rapid Application Development (RAD) model contains the characteristics of iterative development and prototyping model. Functional modules are developed simultaneously as prototypes [22]. These modules are then combined to ample the software. It accommodates the change in requirements. It also tracks and reduces process development time with less people and upsurge the reusability of prototypes. Highly skilled personnel are necessary to analyze the business requirements and development of software. It also demands client involvement throughout the different stages of the model.

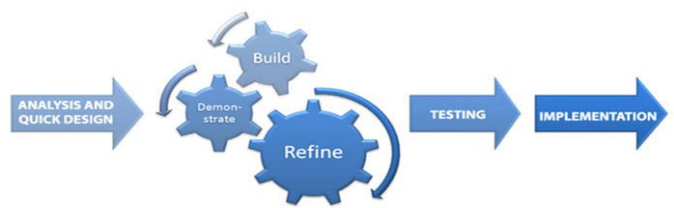

Fig. 6. Rapid Application Development [23]

## III. AGILE MODEL

Agile models are the subgroup of evolutionary models which come with the concept of agility in software engineering [24]. In agile model, the key characteristic is the length of the each iteration. The length of each iteration is two weeks to one month. Each iteration comes up with the outcome of a small version release of the product. Each release is based on functionalities of earlier build versions. Agile models are people-oriented software development life cycle models rather unlike plan driven models. The agile model facilitates with the increased involvement of customers and adapts flexible change in product requirements [25]. Different variations of agile SDLC were being proposed by practitioners and researchers [1]. Basic characteristics of several popular agile models are described below based on their use in industries:

*A. Scrum*

Scrum is widely used agile SDLC model in software industries. Here the life cycle is divided into three main stages: Pre-game, Game/Development, Post-game. The small working team is one of the principles of scrum methodology. Scrum adapts technical and business challenges to provide best products [26]. Scrum consists of sprints which are the iterations from one week to four weeks in length. In scrum, the process is well inspected, changes are adapted, tested as well as documented. Scrum prefer less team members with expertise in agile development. Here requirement selection is prioritized to aid the business value to the clients where well adaption of any addition of requirements which known as a product backlog. Scrum methodology needs team members to be working in same geological location because scrum meetings are needed every day where last identified obstacles, solutions, current product state and future state is discussed. At the end of each sprint, a small version or demo has been released to come up the validation from clients.

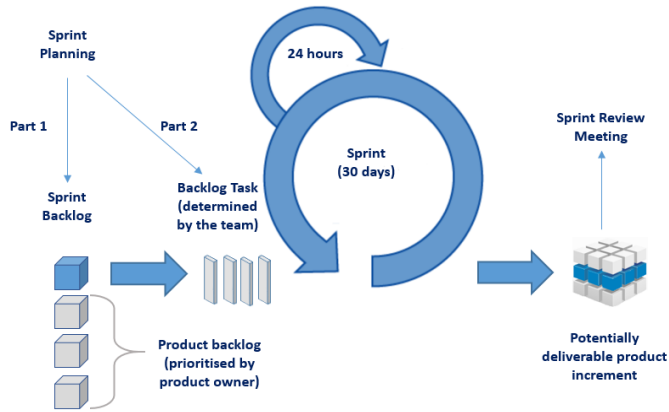

Fig. 7. Scrum Model [27]

*B. Dynamic System Development Method*

Dynamic system development model delivers the systems which encounter tight time by means of incremental prototyping in a meticulous environment [29]. Increments carry adequate functionalities to go forward to the next increment. In this method, practitioners use time boxes to set time and resources which provide the knowledge of the functionalities to be delivered in an increment. DSDM needs active user involvement and the team has given to decide of changes. Changes on requirements are baselined at a welcome meeting. Testing and integration are going on throughout the life cycle.

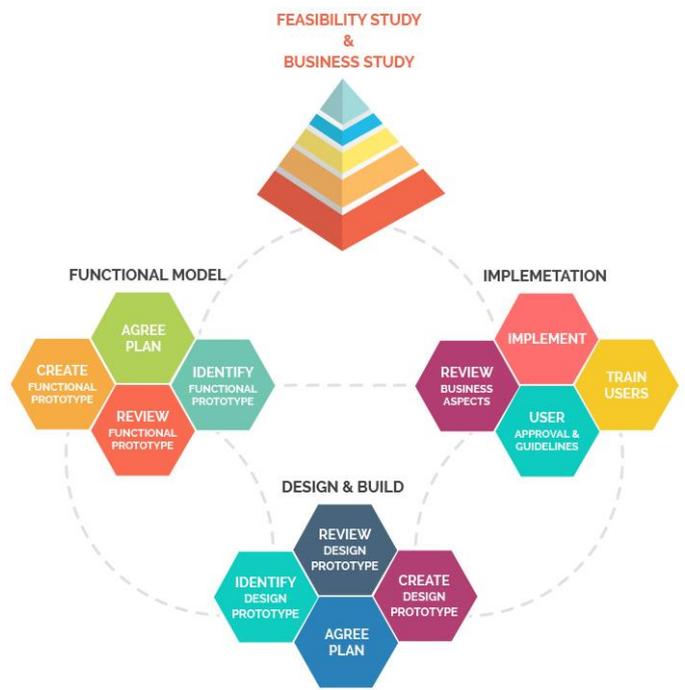

Fig. 8. Dynamic System Development method [28]

*C. Extreme Programming*

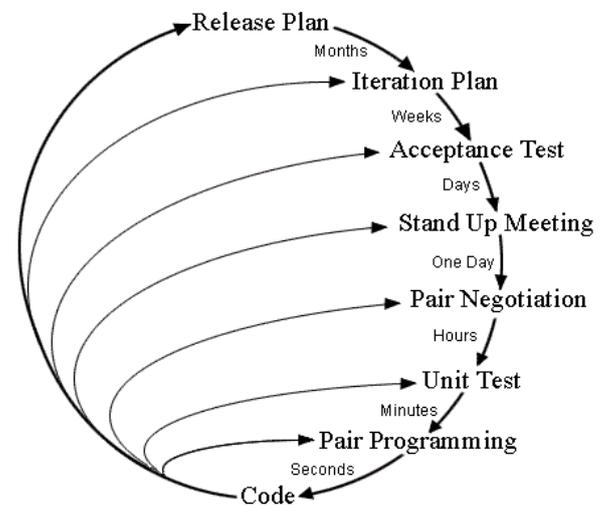

Fig. 9. Extreme Programming [30]

Extreme programming (XP) is established based on object oriented programming. Here focus is on the risks in software development [31]. In planning phase, user stories are created by customer's value. CRC cards and prototypes are the important part of the designing phase in XP. Here pair programming is an important fact for prioritizing the fast development of software. Unit tests are established before initiation of code development and encourage automated testing environment and validation of testing on daily basis. The XP SDLC improves the development in following essential ways: communication, simplicity, daily feedback, respect, courage. XP is suitable for small teams to deal with. XP provides small and frequent version releases. XP provides easy manage through informal methods.

### D. *Feature Driven Model*

Feature Driven Model (FDD) provides the practical object oriented development environment [32]. Using FDD a client-valued feature can be developed in two weeks or less than two weeks. FDD scale down any large product. As it is a feature based model, so for releasing the project features need to defined on early stage and a feature list is prioritized. It needs strong collaborations among team members. Features need to enough size to develop within a short time. FDD is divided into five phases: Developing of an overall model, features list establishing, planning by analyzing the feature, designing by feature and build by feature. FDD is mostly appropriate to develop big products with less consideration to the initial design.

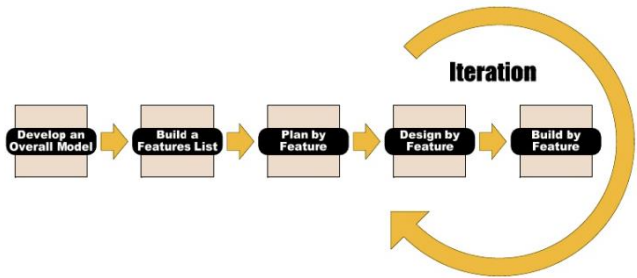

Fig. 10. Feature Driven Model

### E. *Crystal*

Crystal SDLC sets maneuverability for limited resource projects of invention and communication. It has the main goal to bring useful software and a secondary goal is to set up the projects for its next phase. It provides always face-to-face communication with collaborators. For big systems, the teams become larger and the process becomes heavier. It has the principles of skilled, disciplined and understandable encounter of process and documentation. Those team members who are not working on critical phase, may put effort on their extra time for refining the product or helping people who are on the critical phase. Incremental development strategy also used here but the length varied from 1 to 3 months. Currently, this new model has three methodologies: Clear (projects with low critical state), Orange (Moderate critical project), Orange Web (Critical application of e-business).

## IV. COMPARATIVE ANALYSIS

Currently, software industries follow two basic type of software development process models: Plan driven models, Agile Models. Software industries select their development approach based on their product requirements, personnel, team skills, problem complexities, organizational needs, organization size, organizations geolocations etc. In the, following subsections, some basic criteria regarding the selection of SDLC model has been discussed:

### A. *Product Size*

The product itself the crucial factor regarding the selection of any SDLC models. Large systems like driver module for space vehicles, aircraft autopilot system, space telescope modules, brain interfacing application etc. needs to be planned and designed with great effort. More scrutinized project planning is required to make such a big system successful. After launching that system software, it will be very costly to change any requirements. So, for large systems plan driven SDLC models like waterfall [11] are used as next stages cannot be started without appropriate results on previous stages. For small systems like e-commerce, classifieds, management systems, driver tools for personal computers developed following agile methodologies as change is inevitable as the update is a continues process in these contexts.

### B. *Critical Fact*

Software process models are also selected by analyzing critical facts of any systems. Critical systems like plain cockpit module systems are hihgly critical system. Any changes on that system will raise the effort to cope up with the updated system as aeroplane itself is a critical system. So, plan driven systems like waterfall can be effective for these systems. Less critical systems like music player can be updated any time it's failure won't cost hundreds of lives. So, for these types of system development, FDD will be the best selection.

### C. *Development Environment*

The environment of the development is a crucial parameter for software development. As we know, agile models succeed by chaos and merging teams together. So, the members in the development team as well as the customers have the freedom to change the requirements. As fast communication is necessary, team members should be in the same location. On the other hand, plan driven models succeed by following stable environment and proper, but the team can collaborate from different geo locations.

### D. *Customer Interactions*

Agile provides flexibility. So, agile models welcome changes on requirements in the development phase or any iterations. On the other hand, plan driven models welcome client involvement on requirement analysis and product delivery.

### E. *Member Experiences*

Agile methodologies for software development need expert agile personnel to come up with the success. For plan driven models, experience and less experience both type of members collaborates to reach the success peak.

From the discussed analysis facts, we understand that different models are best suited for different aspects. The length of each iteration is lengthier in some models and shorter in other models. A specific product is developed on the dynamic environment in the starting phases of the life cycle of the product as well as the environment may stable at the future phases. We have observed that no process model can accommodate both the stable and dynamic environment. We have also found the research gap for a software process model that can incorporate the human centred design of application software as use in ubiquitous application development [33]. Usage of plan driven models like Waterfall, Iterative, Incremental, RAD are suitable for large, critical systems and stable environment. On the other hand, these models will be costly for the dynamic development environment, changes and maintenance of software. Less user involvement also can cause unsuccessful projects using plan driven models. Agile models like Scrum, XP, DSDM are the best suited models for small and medium systems [34], can adapt changes on any iteration, enable shorter iteration and dynamic environment [6]. But this type of freedom regarding change adaption sometimes cause a delay on product deadline. Migration among methodologies [35] also cost time and money as a team needs to cope up with the transition. So, a model needs to be proposed to tradeoff among the suitability of plan driven and agile models.

## V. CONCLUSION

Software industries use various software process models for developing software with the product and organizational success. A successful development depends on the effective use of the several life cycle stages. Current life cycle models are used based on the project size, reusability of the product, development environment, customer interaction. Plan driven SDLC models are suitable for critical and large projects and it needs to be in a stable environment. On the other hand, Agile models are best fitted on the development of small and less critical systems and facilitate the dynamic environment. Plan driven models are going to be costly when a change in product requirement is inevitable. In contrast, agile models are costly when a change in requirements is too frequently that cause wastage of development time. As different strategies or SDLC are suited for different projects, so adaption of different models for team members at the transition period costs the time. So, delay on delivery is a regular issue [37]. In this research paper, we have discovered and analyzed the facts of SDLC models with respect to software industries. The analyzed facts prove that it is obvious to propose a new SDLC model which tradeoffs among the suitability of plan driven and agile models to improve the outcomes of both clients and the software industries.

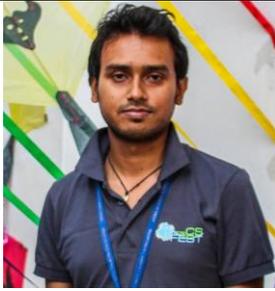
AZM Ehtesham Chowdhury completed B.Sc. in Computer Science & Engineering and M.Sc. in Computer Science from American International University-Bangladesh, Dhaka, Bangladesh. His current research interest includes Data Science, Data Mining, Software Engineering, Intelligent Systems. Computer Vision, Pattern Recognition and Human Computer Interaction.

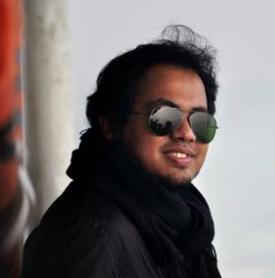
Abhijit Bhowmik completed his B.Sc. in Computer Science & Engineering and M.Sc. in Computer Science from the American International University–Bangladesh (AIUB). Currently he is working as a Senior Assistant Professor and Special Assistant, Office of Student Affairs (OSA) in the Department of Computer Science, AIUB. His research interests include wireless sensor networks, video on demand, software engineering, mobile & multimedia communication and data mining.

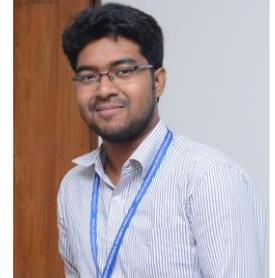
Md Hasibul Hasan obtained his B.Sc. in Computer Science & Software Engineering and M.Sc. in Computer Science from American International University-Bangladesh, Dhaka, Bangladesh. His current research interest includes Data Mining and Software Engineering.

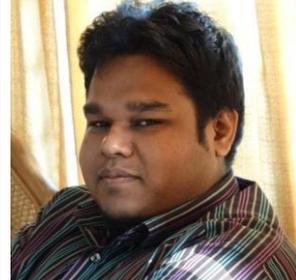
Md Shamsur Rahim obtained his B.Sc. in Computer Science & Software Engineering and M.Sc. in Computer Science from American International University-Bangladesh, Dhaka, Bangladesh. His current research interest includes Data Science, Data Mining and Software Engineering.